\documentclass{JHEP3}

  \usepackage{latexsym,bm,amsmath,amssymb,amsfonts}
  \usepackage{epsfig,graphics,graphicx}
  \usepackage{slashed}
  \usepackage[latin1]{inputenc}
  \usepackage{mathrsfs}
\usepackage{mathcomp}

  \long\def\comment#1{ }

\title{\rm \LARGE \bf Back reaction, covariant anomaly and effective action}

\author{Qing-Quan Jiang$^{a,b}$ and Xu Cai$^{b}$\\

$^{a}$ Institute of Theoretical Physics, China
 West Normal University, Nanchong, Sichuan \\637002, People's Republic of
China\\
$^{b}$ Institute of Particle Physics, Central China Normal
University, Wuhan, Hubei 430079, \\People's Republic of China\\
{\tt E-mail address: jiangqq@iopp.ccnu.edu.cn, xcai@mail.ccnu.edu.cn}}

\abstract{In the presence of back reaction, we first produce the one-loop corrections for the event horizon and Hawking temperature of the Reissner-Nordstr\"{o}m black hole. Then, based on the covariant anomaly cancelation method and the effective action technique, the modified expressions for the fluxes of gauge current and energy momentum tensor, due to the effect of back reaction, are obtained. The results are consistent with the Hawking fluxes of a (1+1)-dimensional blackbody at the temperature with quantum corrections, thus confirming the robustness of the covariant anomaly cancelation method and the effective action technique for black holes with back reaction.}

\begin{document}

\section{Introduction}

Hawking proved that black holes are not perfectly black, rather they can radiate
particles characterized by the thermal spectrum with the
temperature $T=(1/2\pi)\kappa$, where $\kappa$ is the surface
gravity of the black hole\cite{Hawking}. Since then, the study of Hawking
radiation has long been attracted a lot of attentions of
theoretical physicists. Recently, Robinson and Wilczek have proposed a particularly illuminating way to understand Hawking effect via gravitational anomaly\cite{robinson}. In this paper, the effective two dimensional theory is formulated outside the horizon to exclude the classically irrelevant ingoing modes, as a result, gravitational anomaly with respect to diffeomorphism symmetry appears. Unitarity is, of course, preserved, so gravitational anomaly should be cancelled by quantum effects of
the modes that were irrelevant classically. The result shows that the compensating flux to cancel gravitational anomaly at the horizon is equal to that of (1+1)-dimensional blackbody at the Hawking temperature. Later on, the method is widely used to include gauge anomaly for the charged or rotating black holes\cite{canomaly1,add1,canomaly2,canomaly3,canomaly4,canomaly5,canomaly6}.
In these observations\cite{robinson,canomaly1,add1,canomaly2,canomaly3,canomaly4,canomaly5,canomaly6}, the Hawking fluxes are derived by requiring both the \emph{consistent} gauge and gravitational anomalies and the vanishing conditions of \emph{covariant} current and energy momentum tensor. To simplify this consistent anomaly model, Banerjee and Kulkarni proposed a more economical and conceptually cleaner model, called as covariant anomaly, to derive Hawking radiation of black holes\cite{anomaly1}. Here, the Hawking fluxes can equally well be obtainable from
the \emph{covariant} expression only. Further applications of this approach can be found in Refs.\cite{anomaly2,eanomaly}.

Just at the same time, another effective method
to correctly reproduce Hawking fluxes is also proposed to base on the effective action
by using the \emph{covariant} boundary condition and the \emph{covariant} energy-momentum tensor\cite{eanomaly,effective}. This approach is particularly direct and useful since
only exploitation of known structure of effective action at the
horizon is sufficient to determine the Hawking fluxes of energy-momentum tensor.

Based on these developments for the covariant anomaly cancelation method and the effective action technique, some
elaborate applications soon appeared in Refs.\cite{anomaly1,anomaly2,eanomaly,effective}. None of
these computations, however, consider the effect of back reaction. Naturally, it becomes interesting to incorporate the effect of back reaction in these analysis of \cite{anomaly1,anomaly2,eanomaly,effective}. Now a question arises whether the covariant
anomaly cancelation technique and the effective action method,
having been largely proved their robustness, are still applicable to include quantum corrections.

In this paper, starting
from the quantum-corrected event horizon and Hawking temperature, we first adopt the covariant anomaly cancelation method to produce the modified Hawking fluxes including the effect of back reaction. The results are consistent with the Hawking fluxes of a (1+1)-dimensional blackbody at the Hawking temperature with quantum corrections, thus confirming the robustness of the covariant anomaly cancelation method for black hole with back reaction. Then, we apply the effective action technique to reproduce the same result, also confirming its robustness for black holes including the effect of back reaction.

The remainders of this paper is organized as follows. In
Sec.\ref{2}, we first reproduce the Reissner-Nordstr\"{o}m black hole including the effect of back reaction, and provide the one-loop corrections for its event horizon and Hawking temperature. In the presence of back reaction, Sec.\ref{4} and \ref{5} are, respectively, devoted to check the robustness of the covariant anomaly cancelation method and the effective action technique. The modified expressions for the Hawking fluxes, due to the effect of back reaction, are obtained.
Sec.\ref{6} contains some conclusions and discussions.

\section{Back reaction on the Reissner-Nordstr\"{o}m black hole }\label{2}

In this section, we aim to study the effects of back reaction on the Reissner-Nordstr\"{o}m black hole and from there, the one-loop corrections for its event horizon and Hawking temperature are obtained. For the Reissner-Nordstr\"{o}m black hole, the vacuum solution can be written as
\begin{equation}
ds^2=-f(r)dt^2+\frac{1}{f(r)}dr^2+r^2d\theta^2+r^2\sin^2\theta d\phi^2,
\end{equation}
where $f(r)=1-\frac{2m}{r}+\frac{Q^2}{r^2}$, and the gauge potential $\mathcal{A}_t(r)=-\frac{Q}{r}$. The usual event horizon is located at $r=r_h=m+\sqrt{m^2-Q^2}$.
The usual Hawking temperature is given by $T_h\equiv\frac{\kappa_h}{2\pi}\equiv\frac{1}{4\pi}\partial_rf(r)|_{r_h}$. In the semiclassical approximation, the same results for the event horizon and Hawking temperature can also be reproduced by the tunneling formalism. Beyond the semiclassical approximation, the results would experience higher-order quantum corrections(see in Refs.\cite{correction1,correction2,correction3,spectrum}).

The effect of back reaction, acting as a source of curvature\cite{New1}, stems from the non-zero vacuum expectation value of the energy momentum tensor. When the matter fields $(\phi)$ (including the graviton) are quantized at the one-loop level and then coupled to gravity, the Einstein equation can be written as
\begin{equation}
\mathcal{R}_{\mu\nu}-\frac{1}{2}\mathcal{R}g_{\mu\nu}=\langle T_{\mu\nu}(\phi,g_{\mu\nu})\rangle, \label{eq1}
\end{equation}
By solving Eq.(\ref{eq1}), and assuming that far away from the black hole we have essentially vacuum, the Reissner-Nordstr\"{o}m black hole including the effect of back reaction takes the form as\cite{New1}
\begin{equation}
ds^2_{\textrm{corr}}=-\mathcal{F}_{\textrm{corr}}(r)dt^2+\frac{dr^2}{\mathcal{H}_{\textrm{corr}}(r)}+r^2d\theta^2+r^2\sin^2\theta d\phi^2,\label{2eq9}
\end{equation}
where
\begin{eqnarray}
\mathcal{H}_{\textrm{corr}}(r)&=&1-\frac{2m}{r}+\frac{Q^2}{r^2}+\frac{1}{r}\int_{r_\infty}^{r}r^2T_t^tdr, \label{eq7}\nonumber\\
\mathcal{F}_{\textrm{corr}}(r)&=&\mathcal{H}_{\textrm{corr}}e^{\int_{r_\infty}^r(T_r^r-T_t^t)r\mathcal{H}^{-1}_{\textrm{corr}}dr}.\label{3eq9}
\end{eqnarray}
Obviously, the corrected terms in (\ref{3eq9}) are related to the non-zero vacuum
expectation value of the energy momentum tensor. Here, we use the thermal stress-energy tensor given by Huang \cite{New2} for the case of the Reissner-Nordstr\"{o}m black hole. As a result, the corrected metric due to back reaction can be fully determined.

Next, we focus on providing the one-loop corrections for the event horizon and Hawking temperature. After carefully analyzing the dimensions and
the integrability condition for the first law of thermodynamics of the black hole, the most general term which has the
dimension of $\hbar$ can be reformulated as
\begin{equation}
\mathcal{H}_{RN}(m,Q)=a_2\Big(mr_h-\frac{1}{2}Q^2\Big), \label{neq13}
\end{equation}
whose concrete derivation appears in Appendix A. Here, $a_2$ is a dimensionless constant, and $r_h=m+\sqrt{m^2-Q^2}$ is the usual horizon of the Reissner-Nordstr\"{o}m black hole. For the static black hole, the event horizon is determined by solving the equation $g_{tt}(r_+)=0=g^{rr}(r_+)$. Under this condition, the modified horizon radius at the one-loop level can then be written as
 \begin{equation}
 r_+=r_h\Big(1+\beta\frac{\hbar}{\mathcal{H}_{RN}}\Big), \label{eq14}
\end{equation}
where
\begin{equation}
\beta=-\frac{\mathcal{H}_{RN}}{\hbar}\frac{1}{2(m-Q^2/r_h)}\int_{r_\infty}^{r_h}r^2T_t^tdr. \label{eq15}
\end{equation}
is a dimensionless constant parameter.
Here, the dimensionless constant $\beta$ can be fully determined since the integral $\int_{r_\infty}^rr^2T_t^tdr$ is known to us in view of the renormalised energy momentum tensor\cite{New2}. Similarly, the one-loop corrections for the surface gravity of the black hole can be read off
\begin{equation}
\kappa_+=\kappa_h\Big(1+\alpha\frac{\hbar}{\mathcal{H}_{RN}}\Big), \label{eq16}
\end{equation}
where $\alpha$ is a dimensionless constant and $\kappa_h=\frac{m}{r_h^2}-\frac{Q^2}{r_h^3}$ is the surface gravity of the black hole without the presence of back reaction, and $\kappa_+=\frac{1}{2}\partial_r\sqrt{\mathcal{F}_{\textrm{corr}}\mathcal{H}_{\textrm{corr}}}|_{r_+}$ is the modified surface gravity due to back reaction, and
\begin{equation}
\alpha =\frac{\mathcal{H}_{RN}}{\hbar}\Big[\frac{1}{2}\int_{r_\infty}^{r_+}\Big(T_t^t-T_r^r\Big)r \mathcal{H}^{-1}_{\textrm{corr}}dr
+\frac{1}{\kappa_h}\Big(r_+T_t^t(r_+)-\frac{1}{r_+^2}\int_{r_\infty}^{r_+}r^2T_t^tdr\Big)\Big].\label{eq17}
\end{equation}
Here, since the integrals $\int_{r_\infty}^rr^2T_t^tdr$ and $\int_{r_\infty}^r(T_r^r-T_t^t)r\mathcal{H}^{-1}_{\textrm{corr}}dr$ can be completely determined by the renormalised
energy momentum tensor\cite{New2}, the dimensionless parameter $\alpha$ can also be known to us. Thus, the modified Hawking temperature at the one-loop level can be written as
\begin{equation}
T_{\textrm{corr}}=\frac{\kappa_+}{2\pi}=\frac{\kappa_h}{2\pi}\Big(1+\alpha\frac{\hbar}{\mathcal{H}_{RN}}\Big). \label{eq11}
\end{equation}
Till now, as a result of back reaction, we have obtained the one-loop level corrections to the horizon radius and Hawking temperature. It should be noted that $\beta$ is negative and the effect of back reaction will be a decrease in the usual horizon radius. While the Hawking temperature (\ref{eq11}) with the one-loop corrections is a perturbative solution, and the constant $\alpha$ is related to the matter-graviton balance. That is, if $\alpha >0$ (i.e when the matter (scalar field) is dominant), the surface gravity gets enhanced by the effect of back reaction, while for $\alpha <0$ (i.e when the graviton is dominant), it gets reduced by the one-loop back reaction effect\cite{New1}. In fact, the dimensionless constant $\alpha$ is a model-dependent parameter. In the case of Loop Quantum Gravity,
$\alpha$ is a negative coefficient whose exact value was once an object of debate but has since been rigorously fixed at $\alpha=-\frac{1}{2}$. In the string theory, the sign of $\alpha$ depends on the number of field species appearing in the low energy approximation.

In the next sections, basing on the covariant anomaly cancelation method and the effective action technique, we focus on deriving the modified Hawking fluxes of the Reissner-Nordstr\"{o}m black hole with back reaction .

\section{Back reaction and covariant anomalies} \label{4}

As a warm-up, we briefly review the basics idea of applying the covariant anomaly cancelation approach to derive the Hawking fluxes.
An anomaly in a quantum field theory is a conflict between a symmetry of the classical action and the procedure of quantization.
Since the horizon is a null hypersurface, modes interior to the horizon cannot, classically, affect physics outside the horizon. If we
formally remove the classically irrelevant ingoing modes to obtain the effective action in the exterior region, it becomes anomalous
with respect to gauge and diffeomorphism symmetries. The underlying theory is, of course, invariant. To restore gauge invariance
and general coordinate covariance at the quantum level, one must introduce the fluxes of gauge charge current and energy
momentum tensor to cancel gauge and gravitational anomalies at the horizon, which is compatible with the Hawking fluxes of the charge and energy momentum tensor.

In this section, we adopt covariant anomaly cancelation method to derive Hawking radiation of the Reissner-Nordstr\"{o}m black hole with back reaction, and expect that the compensating fluxes required to cancel the covariant gauge and gravitational
anomalies at the horizon have an equivalent form to those of Hawking radiation with a chemical potential. To complete that, near
the black hole horizon, we first introduce a dimensional reduction technique to reduce the higher-dimensional theory to the effective two-dimensional one. For simplicity, we consider the action of a massless scalar field in the black hole background with the minimal electro-magnetic coupling term. Upon introducing the tortoise coordinate transformation $\frac{\partial r_\ast}{\partial r}=\frac{1}{\sqrt{\mathcal{F_{\textrm{corr}}}(r)\mathcal{H}_{\textrm{corr}}(r)}}$ and a partial wave decomposition as $\Psi(t,r,\theta,\phi)=Y_{lm}(\theta,\phi)\Psi(t,r)$, the physics near the horizon can be described by an infinite set of (1+1)-dimensional fields in the spacetime with the metric
\begin{equation}
ds^2=-\mathcal{F_{\textrm{corr}}}(r)dt^2+ \frac{dr^2}{\mathcal{H}_{\textrm{corr}}(r)}, \label{eq36}
\end{equation}
and the gauge potential $\mathcal{A}_t(r)=-\frac{Q}{r}$. In the two-dimensional reduction, when omitting the classically irrelevant ingoing modes at the horizon, the effective theory contains gauge and gravitational anomalies. To restore gauge invariance and diffeomorphism covariance at the quantum level, we should introduce the fluxes of gauge current and energy momentum tensor to cancel gauge and gravitational anomalies. Let's, first, study gauge anomaly and gauge current flux.

As is well-known that the covariant form for the anomalous Ward identity can be written as $\nabla_\mu J^\mu=-\frac{e^2}{4\pi\sqrt{-g}}\epsilon^{\mu\nu}F_{\mu\nu}=\frac{e^2}{2\pi\sqrt{-g}}\partial_r\mathcal{A}_t(r)$.
Near the horizon, the classically irrelevant ingoing modes is excluded to formulate the effective field theory, so
the gauge current here exhibits an anomaly with respect to gauge invariance. For the left-handed field, the covariant gauge
current satisfies $\partial_r[\sqrt{-g}J_{(H)}^r]=\frac{e^2}{2\pi}\partial_r\mathcal{A}_t(r)$. Away from the horizon, since there is no anomaly, the gauge current is conserved here, and satisfies $\nabla_\mu J_{(o)}^\mu=0$. Integrating the gauge currents in the two regions yields
\begin{eqnarray}
\sqrt{-g}J_{(o)}^r&=&c_o, \nonumber\\
\sqrt{-g}J_{(H)}^r&=&c_H+\frac{e^2}{2\pi}\int_{r_+}^rdr \partial_r\mathcal{A}_t(r), \label{eq37}
\end{eqnarray}
where $c_o$ and $c_H$ are integration constants, respectively denoting the covariant gauge currents at $r=\infty$ and $r=r_+$. If introducing two step functions $\Theta_+(r)=\Theta(r-r_+-\varepsilon)$ and $H(r)=1-\Theta_+(r)$, the total current can be written as $J^\mu=J_{(o)}^\mu\Theta_+(r)+J_{(H)}^\mu H(r)$. Then the Ward identity can be rewritten as
\begin{equation}
\partial_r[\sqrt{-g}J^r]=\partial_r\Big(\frac{e^2}{2\pi}\mathcal{A}_t(r)H(r)\Big)+\Big[\sqrt{-g}\big(J_{(o)}^r-J_{(H)}^r\big)+\frac{e^2}{2\pi}
\mathcal{A}_t(r)\Big]\delta(r-r_+-\varepsilon). \label{eq38}
\end{equation}
To make the gauge anomaly free under gauge transformation, the first term should be canceled by the quantum effect of the classically irrelevant ingoing modes, and the second term should vanish at the horizon, which yields
\begin{equation}
c_o=c_H-\frac{e^2}{2\pi}\mathcal{A}_t(r_+).\label{eq39}
\end{equation}
Considering the regularity requirement of the physical quantities at the further horizon, we impose the boundary condition that the covariant current vanishes at the horizon, which means $c_H=0$. This determines the gauge current flux to be
\begin{equation}
c_o=-\frac{e^2}{2\pi}\mathcal{A}_t(r_+)=\frac{e^2 Q}{2\pi r_h}\Big(1+\beta\frac{\hbar}{\mathcal{H}_{RN}}\Big)^{-1}.\label{eq40}
\end{equation}
This is the modified gauge current flux in the presence of back reaction, whose value is expected to equal that of a (1+1)-dimensional blackbody at the temperature with quantum corrections, as will appear presently. When $\hbar\rightarrow 0$, we can reproduce the usual current flux as Ref.\cite{canomaly1}.

Apart from gauge anomaly, when excluding the classically irrelevant ingoing modes at the horizon, gravitational anomaly also takes place as a consequence as a breakdown of the  diffeomorphism symmetry, which is normally expressed as the nonconservation of the energy momentum tensor. Away from the horizon, there is no anomaly, but contains an effective background gauge potential, so the energy momentum tensor satisfies the modified conservation equation as $\nabla_\mu T_{(o)\nu}^\mu=F_{\mu\nu}J_{(o)}^\mu$. While the energy momentum tensor near the horizon obeys the anomalous Ward identity after adding
gravitational anomaly as $\nabla_\mu T_{(H)\nu}^\mu=F_{\mu\nu}J_{(H)}^\mu+A_\nu$, where $A_\nu=-\frac{1}{96\pi}\sqrt{-g}\epsilon_{\mu\nu}\partial^\mu R$. For the metric of the form (\ref{eq36}), the anomaly is purely timelike, which means $A_r=0$ and $A_t=\frac{1}{\sqrt{-g}}\partial_rN_t^r(r)$ where $N_t^r(r)=\frac{1}{96\pi}\big(\mathcal{H}_{\textrm{corr}}\mathcal{F}''_{\textrm{corr}}+\frac{1}{2}\mathcal{H}'_{\textrm{corr}}\mathcal{F}'_{\textrm{corr}}
-\frac{1}{\mathcal{F}_{\textrm{corr}}}\mathcal{F}'^2_{\textrm{corr}}\mathcal{H}_{\textrm{corr}}\big)$.  Now, integrating the energy momentum tensor in both regions, we have
\begin{eqnarray}
\sqrt{-g}T_{(o)t}^r&=&a_o+c_o\mathcal{A}_t(r),\nonumber\\
\sqrt{-g}T_{(H)t}^r&=&a_H+\int_{r_+}^rdr\partial_r\big(c_o\mathcal{A}_t(r)+\frac{e^2}{4\pi}\mathcal{A}_t^2(r)+N_t^r(r)\big),\label{eq41}
\end{eqnarray}
where $a_o$ and $a_H$ are integration constants, respectively representing the values of the energy flow at the infinity and the horizon.
Writing the total energy momentum tensor as $T_\nu^\mu=T_{(o)\nu}^\mu\Theta_+(r)+T_{(H)\nu}^\mu H(r)$, we then have
\begin{eqnarray}
\sqrt{-g}\nabla_\mu T_t^\mu &=&c_o\partial_r\mathcal{A}_t(r)+\partial_r\big[\big(\frac{e^2}{4\pi}\mathcal{A}_t^r(r)+N_t^r\big)H(r)\big]\nonumber\\
&+&\big[\sqrt{-g}\big(T_{(o)t}^r-T_{(H)t}^r\big)+\frac{1}{4\pi}\mathcal{A}_t^2(r)+N_t^r(r)\big]\delta(r-r_+-\varepsilon).\label{eq42}
\end{eqnarray}
In Eq.(\ref{eq42}), the first term is the classical effect of the background electric field for constant current flow. The second term should be canceled by the quantum effect of the classically irrelevant ingoing modes. In order to restore the diffeomorphism covariance at the quantum level, the third term must vanish at the horizon, which yields
\begin{equation}
a_o=a_H+\frac{e^2}{4\pi}\mathcal{A}_t^2(r_+)-N_t^r(r_+).\label{eq43}
\end{equation}
To fix $a_o$ completely, it is necessary to impose the boundary condition that the covariant energy momentum tensor vanishes at the horizon, which means $a_H=0$. So the total flux of the energy momentum tensor is given by
\begin{eqnarray}
a_o&=&\frac{e^2}{4\pi}\mathcal{A}^2_t(r_+)-N_t^r(r_+)=\frac{e^2}{4\pi}\mathcal{A}_t^2(r_+)+\frac{1}{192\pi}\mathcal{F}'_{\textrm{corr}}(r_+)\mathcal{H}'_{\textrm{corr}}(r_+),\nonumber\\
&=&\frac{e^2}{4\pi}\mathcal{A}^2_t(r_+)+\frac{\pi}{12}T_{\textrm{corr}}^2\nonumber\\
&=&\frac{e^2Q^2}{4\pi r_h^2}\Big(1+\beta\frac{\hbar}{\mathcal{H}_{RN}}\Big)^{-2}+\frac{\pi}{12}T^2_h\Big(1+\alpha\frac{\hbar}{\mathcal{H}_{RN}}\Big)^{2}.\label{eq44}
\end{eqnarray}
This is the compensating energy flux for canceling the covariant anomaly at the horizon, whose value is expectantly equal to that of a (1+1)-dimensional blackbody with the temperature undergoing quantum corrections. When setting $\hbar\rightarrow 0$, the usual energy flux, as appears in Ref.\cite{canomaly1}, can also be reproduced.

Next, we will produce the fluxes from blackbody radiation at the temperature $T_{\textrm{corr}}$ with a chemical potential, so as to compare the results with the compensating fluxes for canceling gauge and gravitational anomalies at the horizon. The Planck distribution for the Reissner-Nordstr\"{o}m black hole with back reaction can be written as $\mathcal{N}_{\pm e} (\omega)=\frac{1}{e^{(\omega\pm e\mathcal{A}_t(r_+))/T_{\textrm{corr}}}\pm 1}$ for bosons and fermions, respectively. Here, we only consider the fermion case without the loss of generality. With the Planck distribution, the fluxes of charge and energy momentum tensor are
\begin{eqnarray}
(\textrm{Flux})_{\textrm{charge}}&=&e\int_0^\infty\frac{d\omega}{2\pi}[\mathcal{N}_{ e} (\omega)
-\mathcal{N}_{- e} (\omega)]\nonumber\\
&=&-\frac{e^2}{2\pi}\mathcal{A}_t(r_+)=\frac{e^2 Q}{2\pi r_h}\Big(1+\beta\frac{\hbar}{\mathcal{H}_{RN}}\Big)^{-1},\label{eq34}\\
(\textrm{Flux})_{\textrm{energy}}&=& \int_0^\infty\frac{d\omega}{2\pi}\omega[\mathcal{N}_{ e} (\omega)+\mathcal{N}_{ -e} (\omega)]\nonumber\\
&=&\frac{e^2}{4\pi}\mathcal{A}_t^2(r_+)+\frac{\pi}{12}T_{\textrm{corr}}^2\nonumber\\
&=&\frac{e^2Q^2}{4\pi r_h^2}\Big(1+\beta\frac{\hbar}{\mathcal{H}_{RN}}\Big)^{-2}+\frac{\pi}{12}T^2_h\Big(1+\alpha\frac{\hbar}{\mathcal{H}_{RN}}\Big)^{2}. \label{eq35}
\end{eqnarray}

 Obviously, the results (\ref{eq40}) and (\ref{eq44}) derived from the covariant anomaly cancelation approach are compatible with the results (\ref{eq34}) and (\ref{eq35}). So we can easily find that, even for the black hole with back reaction, the thermal fluxes required by black hole thermodynamics are still capable of canceling the covariant gauge and gravitational anomalies at the horizon, and restoring gauge invariance and general coordinate covariance at the quantum level to hold in the effective theory. Thus we have successfully verified the robustness of the covariant anomaly cancelation method in the presence of back reaction. Based on the same idea, in the next section, we aim to check whether the effective action technique is still applicable to include back reaction.

\section{Back reaction and effective action}\label{5}

In this section, we will base solely on the structure of the effective action and the covariant boundary conditions at the horizon to
cross-check thermodynamic properties of the black hole with back reaction. As mentioned in Sec.\ref{4}, with the aid of the dimensional reduction technique, the higher theory near the horizon can be effectively described by the two dimensional chiral theory. For the reduced two-dimensional chiral theory, the expression for the anomalous (chiral) effective action has already been derived \cite{effective,New5}. Now taking appropriate functional derivatives of the anomalous effective action yields the anomalous gauge current and energy momentum tensor near the horizon. As a result, some parameters appears to be determined. To accomplish it, we then impose the boundary condition that the covariant gauge current and energy momentum tensor vanish at the horizon. Once these are fixed, the fluxes of the charge and energy momentum tensor can be obtained by taking the asymptotic ($r\rightarrow \infty$) limits of the anomalous gauge current and energy momentum tensor.

Near the horizon, the chiral effective action can be written as \cite{New5}
\begin{equation}
\Gamma_{(H)}=-\frac{1}{3}z(\omega)+z(\mathcal{A}), \label{eq45}
\end{equation}
with
\begin{equation}
z(v)=\frac{1}{4\pi}\int d^2xd^2y\epsilon^{\mu\nu}\partial_\mu v_\nu(x)\Box^{-1}(x,y)\partial_\rho[(\epsilon^{\rho\sigma}+\sqrt{-g}g^{\rho\sigma})v_\sigma(y)], \label{eq46}
\end{equation}
where $\Box=\nabla^\mu\nabla_\mu$ is the Laplacian operator in the two-dimensional background, and $\mathcal{A}_\mu$ and $\omega_\mu$ are, respectively, the gauge field and the spin connection. Carrying on a variation for the effective action, we can obtain the gauge current and energy momentum tensor. To get their covariant forms in which we are interested, one needs to add appropriate local polynomials \cite{New5}. The final expressions for the covariant gauge current $J^\mu$ and energy momentum tensor $T_\nu^\mu$ can be obtained as
\begin{eqnarray}
J^\mu &=& -\frac{e^2}{2\pi}D^\mu B, \label{eq47}\\
T_\nu^\mu &=& \frac{e^2}{4\pi}D^\mu BD_\nu B+\frac{1}{96\pi}\big(\frac{1}{2}D^\mu FD_\nu F-D^\mu D_\nu F+\delta_\nu^\mu R\big), \label{eq48}
\end{eqnarray}
where $D_\mu=\nabla_\mu-\sqrt{-g}\epsilon_{\mu\nu}\nabla^\nu=-\sqrt{-g}\epsilon_{\mu\nu}D^\nu$ is the chiral covariant derivative. It should be noted that the anomalous Ward identities can be obtained by taking the covariant divergence of (\ref{eq47}) and (\ref{eq48}). Here, the definitions of $B(x)$ and $F(x)$ are provided by
\begin{eqnarray}
B(x)&=&\int d^2y\sqrt{-g}\Box^{-1}(x,y)\epsilon^{\mu\nu}\partial_\mu \mathcal{A}_\nu(y),\nonumber\\
F(x)&=& \int d^2y \Box^{-1}(x,y)\sqrt{-g}R(y). \label{eq49}
\end{eqnarray}
which satisfy the equations $\Box B=-\partial_r\mathcal{A}_t(r)$ and $\Box F=R$. Solving the two equations yield $B=B_o(r)-at+b$ and $F=F_o(r)-4pt+q$, where $B_o(r)$ and $F_o(r)$ satisfy $\partial_rB_o(r)=\frac{1}{\sqrt{\mathcal{F}_{\textrm{corr}}\mathcal{H}_{\textrm{corr}}}}(\mathcal{A}_t(r)+c)$ and $\partial_rF_o(r)=\frac{1}{\sqrt{\mathcal{F}_{\textrm{corr}}\mathcal{H}_{\textrm{corr}}}}(\sqrt{\frac{\mathcal{H}_{\textrm{corr}}}{\mathcal{F}_{\textrm{corr}}}}\mathcal{F}'_{\textrm{corr}}+z)$, in which $a,b,c,p,q,z$ are constants of integration.

Next, based on the expressions for the anomalous gauge current (\ref{eq47}) and energy momentum tensor (\ref{eq48}), we focus on calculating the fluxes of gauge charge and energy momentum tensor. First we investigate the gauge charge flux. Considering the $\mu=r$ component of the anomalous covariant gauge current $J^\mu$, we have
\begin{equation}
J^r(r)=\frac{e^2}{2\pi}\sqrt{\frac{\mathcal{H}_{\textrm{corr}}}{\mathcal{F}_{\textrm{corr}}}}\big[\mathcal{A}_t(r)+a+c\big]. \label{eq50}
\end{equation}
To fix the parameters in Eq.(\ref{eq50}), we impose the boundary condition that the covariant gauge current vanishes at the horizon, which means $J^r(r_+)=0$. As a result, $a+c=-\mathcal{A}_t(r_+)$. As it is well-known, the charge flux is often determined by the asymptotic ($r\rightarrow\infty$)
limits of the covariant anomalous free current. In fact, in this limit, we observe that the anomaly vanishes. So the charge flux can be directly obtained by taking the asymptotic limit of Eq.(\ref{eq50}) multiplied by an overall factor of $\sqrt{-g}$, which yields
\begin{equation}
c_o=(\sqrt{-g}J^r)(r\rightarrow\infty)=-\frac{e^2}{2\pi}\mathcal{A}_t(r_+)=\frac{e^2 Q}{2\pi r_h}\Big(1+\beta\frac{\hbar}{\mathcal{H}_{RN}}\Big)^{-1}.\label{eq51}
\end{equation}
This is the modified charge flux based on the effective action technique, whose value exactly agrees with the charge flux (\ref{eq34}). So, in the presence of back reaction, the effective action technique can correctly reproduce the modified charge flux of Hawking radiation.

Now we focus our attention on the flux of energy momentum tensor. The $r-t$ component of the anomalous covariant energy momentum tensor (\ref{eq48}) can be derived as
\begin{eqnarray}
T_t^r(r)&=&\frac{e^2}{4\pi}\sqrt{\frac{\mathcal{H}_{\textrm{corr}}}{\mathcal{F}_{\textrm{corr}}}}\big[\mathcal{A}_t(r)-\mathcal{A}_t(r_+)\big]^2\nonumber\\
&+&\frac{1}{192\pi}\sqrt{\frac{\mathcal{H}_{\textrm{corr}}}{\mathcal{F}_{\textrm{corr}}}}\big[4p+
\sqrt{\frac{\mathcal{H}_{\textrm{corr}}}{\mathcal{F}_{\textrm{corr}}}}\mathcal{F}'_{\textrm{corr}}+z\big]^2\nonumber\\
&-&\frac{1}{96\pi}\sqrt{\frac{\mathcal{H}_{\textrm{corr}}}{\mathcal{F}_{\textrm{corr}}}}\big[\sqrt{\frac{\mathcal{H}_{\textrm{corr}}}{\mathcal{F}_{\textrm{corr}}}}
\mathcal{F}'_{\textrm{corr}}\big(4p+\sqrt{\frac{\mathcal{H}_{\textrm{corr}}}{\mathcal{F}_{\textrm{corr}}}}
\mathcal{F}'_{\textrm{corr}}+z\big)\nonumber\\
&+&\mathcal{H}_{\textrm{corr}} \mathcal{F}''_{\textrm{corr}}
-\frac{1}{2}\mathcal{F}'_{\textrm{corr}}(\frac{\mathcal{H}_{\textrm{corr}}}{\mathcal{F}_{\textrm{corr}}}\mathcal{F}'_{\textrm{corr}}-\mathcal{H}'_{\textrm{corr}})\big].\label{eq52}
\end{eqnarray}
At the horizon, implementing the boundary condition that the covariant momentum tensor vanishes at the horizon, namely $T_t^r(r_+)=0$, yields the equation $4p=z\pm \sqrt{\mathcal{F}'_{\textrm{corr}}(r_+)\mathcal{H}'_{\textrm{corr}}(r_+)}$. So the total flux of the energy momentum tensor, which is often given by the asymptotic limit for the anomaly free energy momentum tensor, can be given by
\begin{eqnarray}
a_o=(\sqrt{-g}T_t^r)(r\rightarrow\infty)&=&\frac{e^2}{4\pi}\mathcal{A}_t^2(r_+)+\frac{1}{192\pi}\mathcal{F}'_{\textrm{corr}}(r_+)\mathcal{H}'_{\textrm{corr}}(r_+)\nonumber\\
&=&\frac{e^2}{4\pi}\mathcal{A}^2_t(r_+)+\frac{\pi}{12}T_{\textrm{corr}}^2\nonumber\\
&=&\frac{e^2Q^2}{4\pi r_h^2}\Big(1+\beta\frac{\hbar}{\mathcal{H}_{RN}}\Big)^{-2}+\frac{\pi}{12}T^2_h\Big(1+\alpha\frac{\hbar}{\mathcal{H}_{RN}}\Big)^{2}. \label{eq53}
\end{eqnarray}
This is the modified flux of energy momentum tensor obtained from the effective action technique, whose value has the same form as that in Eqs.(\ref{eq35}). So, in the presence of back reaction, the effective action technique can also correctly reproduce the modified energy flux of Hawking radiation. From Eqs.(\ref{eq51}) and (\ref{eq53}), we can conclude that the effective action technique is still applicable for the black hole with back reaction.

\section{Conclusions and discussions}\label{6}

In this paper, our motivation is to check the robustness of the covariant anomaly cancelation method and the effective action technique in the presence of back reaction. To complete that, we first obtain the modified Reissner-Nordstr\"{o}m black hole which captures the effect of back reaction. At the one-loop level, we also produce the modified expressions for its event horizon and Hawking temperature. Then, in the presence of back reaction, we adopt the covariant anomaly cancelation method and the effective action technique to obtain the modified fluxes of charge and energy momentum tensor. The results are completely consistent with the Hawking fluxes of a (1+1)-dimensional blackbody at the temperature with quantum corrections, thus confirming the robustness of the covariant anomaly cancelation method and the effective action technique for black holes with back reaction. When omitting the effect of back reaction (namely, $\hbar\rightarrow 0$), we can also reproduce the same results as appeared in Ref.\cite{canomaly1}.

With the aid of the corrected Hawking temperature (\ref{eq11}), we can proceed with the calculation of the modified Bekenstein-Hawking entropy.
The modified form of the first law of thermodynamics for the Reissner-Nordstr\"{o}m black hole with back reaction can be written as
\begin{equation}
dS_{\textrm{bh}}=\frac{1}{T_{\textrm{corr}}}dm-\frac{\mathcal{A}_t(r_h)}{T_{\textrm{corr}}}dQ.\label{eqn1}
\end{equation}
On the premise that the entropy is a state function for the spacetime even in the presence of back reaction, then $dS_{\textrm{bh}}$ is an exact differential. This yields the following relation
\begin{equation}
\frac{\partial}{\partial m}\Big(-\frac{\mathcal{A}_t(r_h)}{T_{\textrm{corr}}}\Big)\Big|_Q=
\frac{\partial}{\partial Q}\Big(\frac{1}{T_{\textrm{corr}}}\Big)\Big|_m. \label{eqn2}
\end{equation}
If the condition (\ref{eqn2}) holds, the solution of (\ref{eqn1}) can be written as
\begin{equation}
S_{\textrm{bh}}=\int\frac{d m}{T_{\textrm{corr}}}-\int\frac{\mathcal{A}_t(r_h)}{T_{\textrm{corr}}}dQ-\int\frac{\partial}{\partial Q}\Big(\int\frac{d m}{T_{\textrm{corr}}}\Big)dQ.\label{eqn3}
\end{equation}
Let's first solve the integration over $m$, which yields
\begin{eqnarray}
\int\frac{d m}{T_{\textrm{corr}}}  &=&\int\frac{1}{T_h}\big(1+\sum_i(-\alpha)^i\frac{\hbar^i}{\mathcal{H}_{RN}^i}\big)d m\nonumber\\
&=& \frac{\pi}{\hbar}\big(2mr_h-Q^2\big)-2\pi\widehat{\alpha}\log\big(2mr_h-Q^2\big)\nonumber\\
&-&\frac{4\pi\widehat{\alpha}^2\hbar}{2mr_h-Q^2}+\textrm{constant}+\textrm{higher order terms}.\label{eqn4}
\end{eqnarray}
where $\widehat{\alpha}=\frac{\alpha}{a_2}$. In view of the result (\ref{eqn4}), we can easily check the following relation
\begin{equation}
\frac{\partial}{\partial Q}\int\frac{d m}{T_{\textrm{corr}}}=-\frac{\mathcal{A}_t(r_h)}{T_\textrm{corr}}. \label{eqn5}
\end{equation}
Under this condition (\ref{eqn5}), the entropy of the black hole in presence of back reaction is now given by
\begin{equation}
S_{bh}=S_{BH}-2\pi \widehat{\alpha}\log S_{BH}-\frac{4\pi^2\widehat{\alpha}^2}{S_{BH}}+\textrm{constant}+\textrm{higher order terms}, \label{eqn6}
\end{equation}
where $S_{BH}=\frac{\pi}{\hbar}(2mr_h-Q^2)$ is the usual Bekenstein-Hawking entropy for the black hole, and the other terms appear as a result of the presence of back reaction. Obviously, the leading correction is logarithmic while the sub-leading terms involve inverse powers of the entropy. The similar corrections have
been previously presented in field theory methods \cite{other1}, quantum geometry techniques \cite{other2}, general statistical mechanical arguments \cite{other3} and Cardy formula \cite{other4} etc. Here, the dimensionless constant $\widehat{\alpha}$ is related to the trace anomaly as\cite{correction3}
\begin{equation}
\widehat{\alpha}=-\frac{1}{4\pi}\textrm{Im} \int d^4x\sqrt{-g}T_\mu^\mu=\frac{1}{180\pi}\Big(1+\frac{3}{10}\frac{2m^2-r_h^2}{mr_h-Q^2}\Big).
\end{equation}

\appendix

\section{Dimensional analysis}

In the (3+1) dimensions, the Plank length $l_p=\sqrt{\frac{\hbar G}{c^3}}$, the Plank mass $m_p=\sqrt{\frac{\hbar c}{G}}$, and the Plank charge $Q_p=\sqrt{4\pi c \hbar \epsilon_0}$. So, in the unit of $G=c=\frac{1}{4\pi \epsilon_0}$, the dimensions of the Plank length, mass and charge all take the same as $\sqrt{\hbar}$. So, the general form for the dimension of $\hbar$ should be constructed, in terms of black hole parameters, as
\begin{equation}
\mathcal{H}_{RN}(m,Q)=a_1r_h^2+a_2mr_h+a_3m^2+a_4r_hQ+a_5mQ+a_6Q^2,\label{eq54}
\end{equation}
where $a_1, a_2, a_3, a_4, a_5, a_6$ are constants, and $r_h=m+\sqrt{m^2-Q^2}$ is the usual horizon of the Reissner-Nordstr\"{o}m black hole. Noted that the entropy is a state function for all stationary spacetimes even in the presence of quantum correction, which means $dS_{bh}$ must be an exact differential. This condition yields Eq.(\ref{eqn2}), and then leads to
\begin{equation}
\frac{\partial \mathcal{H}_{RN}}{\partial m}\Big|_{Q}=-\frac{1}{\mathcal{A}_t(r_h)}\frac{\partial \mathcal{H}_{RN}}{\partial Q}\Big|_{m}.\label{eq55}
\end{equation}
Substituting Eq.(\ref{eq54}) into Eq.(\ref{eq55}), we have
\begin{eqnarray}
&&(2a_1r_h+a_2m+a_4Q)\Big(1+\frac{m}{\sqrt{m^2-Q^2}}\Big)+a_4r_h+2a_3m+a_5Q \nonumber\\
&&=\frac{r_h}{Q}\Big[(2a_1r_h+a_2m+a_4Q)\frac{Q}{\sqrt{m^2-Q^2}}-a_4r_h-a_5m-2a_6Q\Big]. \label{eq56}
\end{eqnarray}
In Eq.(\ref{eq54}), there are six undetermined constants, so, to fully determine $\mathcal{H}_{RN}$, we need five separate equations but only one equation (\ref{eq55}) is applicable. How can finish that. Supposing the thermodynamic entities at each process for the charged black hole evolving from the non-extremal one to the extremal one ($m=Q$) satisfy the modified first law thermodynamics (\ref{eqn1}), only taking different relation between the mass $m$ and the charge $Q$. Now, we take three cases:

For I case, when $Q=\frac{\sqrt{3}}{2}m$, Eq.(\ref{eq56}) can be rewritten as
\begin{equation}
\frac{3}{2}a_2+2a_3+\frac{\sqrt{3}}{2}a_5=-\frac{3\sqrt{3}}{2}a_4-\sqrt{3}a_5-3a_6.\label{eq61}
\end{equation}

For II case, when $Q=\frac{\sqrt{15}}{4}m$, Eq.(\ref{eq56}) can be read off
\begin{equation}
\frac{5}{4}a_2+2a_3+\frac{\sqrt{15}}{4}a_5=-\frac{5\sqrt{15}}{12}a_4-\frac{\sqrt{15}}{3}a_5-\frac{5}{2}a_6.\label{eq62}
\end{equation}

For III case, when $Q=\frac{2\sqrt{2}}{3}m$, Eq.(\ref{eq56}) is changed as
\begin{equation}
 \frac{4}{3}a_2+2a_3+\frac{2\sqrt{2}}{3}a_5=-\frac{4\sqrt{2}}{3}a_4-\sqrt{2}a_5-\frac{8}{3}a_6.\label{eq63}
\end{equation}

Combining Eqs.(\ref{eq61}), (\ref{eq62}) and (\ref{eq63}), we easily find $a_3=a_4=a_5=0$ and $a_6=-\frac{1}{2}a_2$. So, Eq.(\ref{eq54}) is rewritten as
\begin{equation}
\mathcal{H}_{RN}(m,Q)=a_1r_h^2+a_2mr_h-\frac{1}{2}a_2Q^2.\label{eq64}
\end{equation}
Now, reducing the Reissner-Nodstr\"{o}m black hole to the Schwarzschild black hole (namely, taking $Q=0$) yields
\begin{equation}
\mathcal{H}_{RN}(m,Q)=4a_1m^2+2a_2m^2.\label{eq65}
\end{equation}
From Eq.(\ref{eq65}), we can easily find both the dimensions corresponding to the constants $a_1$ and $a_2$ is the square of the mass $m$. Choosing $a_1=0$, we can obtain Eq.(\ref{neq13}).

\section*{Acknowledgements}
This work is supported by the Natural Science Foundation of
China with Grant Nos. 10773008, 10975062, 70571027, 10635020, a grant by the
Ministry of Education of China under Grant No. 306022.


\begin{thebibliography}{99}

\bibitem{Hawking}
S.W. Hawking, \emph{Particle creation by black holes,
Commun. Math. Phys.} {\bf 43} (1975) 199.

\bibitem{robinson}
S.P. Robinson and F. Wilczek, \emph{Relationship between Hawking Radiation and Gravitational Anomalies, Phys. Rev. Lett.} {\bf 95} (2005) 011303.

\bibitem{canomaly1}
S. Iso, H. Umetsu and F. Wilczek,\emph{Hawking Radiation from Charged Black Holes via Gauge and Gravitational Anomalies, Phys. Rev. Lett.} {\bf 96} (2006) 151302.

\bibitem{add1}
S. Iso, H. Umetsu and F. Wilczek, \emph{Anomalies, Hawking radiations and regularity in rotating black holes, Phys. Rev.} {\bf D74} (2006) 044017;

S. Iso, \emph{Hawking Radiation, Gravitational Anomaly and Conformal Symmetry: The Origin of Universality,
Int. J. Mod. Phys.} {\bf A23} (2008) 2082;

S. Iso, T. Morita and H. Umetsu, \emph{Quantum Anomalies at Horizon and Hawking Radiations in Myers-Perry Black Holes, JHEP.} {\bf 04} (2007) 068;

T. Morita, \emph{Modification of gravitational anomaly method in Hawking radiation, Phys. Lett.} {\bf B677} (2009) 88; \emph{Hawking radiation and quantum anomaly in AdS2/CFT1 correspondence, JHEP.} {\bf 0901} (2009) 037;

K. Umetsu, \emph{Ward Identities in the derivation of Hawking radiation from Anomalies, Prog. Theor. Phys.} {\bf 119} (2008) 849;

K. Murata and U. Miyamoto, \emph{Hawking radiation of a vector field and gravitational anomalies, Phys. Rev.} {\bf D76} (2007) 084038;

U. Miyamoto and K. Murata, \emph{On Hawking radiation from black rings, Phys. Rev.} {\bf D77} (2008) 024020;

K. Murata and J. Soda, \emph{Hawking radiation from rotating black holes and gravitational anomalies, Phys. Rev.} {\bf D74} (2006) 044018.

\bibitem{canomaly2}
H. Shin and W. Kim, \emph{Hawking radiation from non-extremal D1-D5 black hole via anomalies, JHEP.} {\bf 0706} (2007) 012;

W. Kim and H. Shin, \emph{Anomaly Analysis of Hawking Radiation from Acoustic Black Hole, JHEP.} {\bf 07} (2007) 070;

W. Kim, H. Shin and M. Yoon, \emph{Anomaly and Hawking radiation from regular black holes,}  arXiv:0803.3849 [gr-qc];

M. R. Setare, \emph{Gauge and gravitational anomalies and Hawking radiation of rotating BTZ black holes,
Eur. Phys. J.} {\bf C49} (2007) 865;

E. C. Vagenas and S. Das, \emph{Gravitational anomalies, Hawking radiation, and spherically symmetric black holes,
JHEP.} {\bf 10} (2006) 025;

S. Das, S. P. Robinson and E. C. Vagenas, \emph{Gravitational anomalies: A Recipe for Hawking radiation, Int. J. Mod. Phys.} {\bf D17} (2008) 533.


\bibitem{canomaly3}
Q. Q. Jiang, S. Q. Wu and X. Cai, \emph{Anomalies and de Sitter radiation from the generic black holes in de Sitter spaces, Phys. Lett.} {\bf B651} (2007) 65;  \emph{Hawking radiation from the dilatonic black holes via anomalies, Phys. Rev.} {\bf D75} (2007) 064029; \emph{Hawking radiation from the (2+1)-dimensional BTZ black holes, Phys. Lett.} {\bf B651} (2007) 58;

Q. Q. Jiang, \emph{Hawking radiation from black holes in de Sitter spaces,
Class. Quant. Grav.} {\bf 24} (2007) 4391;

 Q. Q. Jiang and S. Q. Wu, \emph{Hawking radiation from rotating black holes in anti-de Sitter spaces via gauge and gravitational anomalies, Phys. Lett.} {\bf B647} (2007) 200;

 S. Q. Wu and J. J. Peng, \emph{Anomalies and Hawking radiation from the Reissner-Nordstrom black hole with a global monopole,
Class. Quant. Grav.} {\bf 24} (2007) 5123.

\bibitem{canomaly4}
X. Kui, W. B. Liu and H. B. Zhang, \emph{Anomalies of the Achucarro-Ortiz black hole,
Phys. Lett.} {\bf B647} (2007) 482;

Z. Xu and B. Chen, \emph{Hawking radiation from general Kerr-(anti)de Sitter black holes, Phys. Rev.} {\bf D75} (2007) 024041;

C. G. Huang, J. R. Su, X. N. Wu and
H. Q. Zhang, \emph{Gravitational Anomaly and Hawking Radiation of Brane World Black Holes, Mod. Phys. Lett.} {\bf A23} (2008) 2957;

X. N. Wu, C. G. Huang and J. R. Sun, \emph{On Gravitational anomaly and Hawking radiation near weakly isolated horizon, Phys. Rev.} {\bf D77} (2008) 124023;

B. Chen and W. He, \emph{Hawking radiation of black rings from anomalies, Class. Quant. Grav.} {\bf 25} (2008) 135011;

J. J. Peng and S. Q. Wu, \emph{Hawking radiation from the Schwarzschild black hole with a global monopole via gravitational anomaly, Chin. Phys.} {\bf B17} (2008) 825;

R. Li, J. R. Ren and S. W. Wei, \emph{Gravitational anomaly and Hawking radiation of apparent horizon in FRW universe, Eur. Phys. J.} {\bf C62} (2009) 455;

S. W. Wei, R. Li and Y. X. Liu and
J. R. Ren, \emph{Anomaly analysis of Hawking radiation from Kaluza-Klein black hole with squashed horizon, Eur. Phys. J.} {\bf C65} (2010) 281;

S. W. Wei, R. Li, Y. X. Liu and J. R. Ren, \emph{Anomaly analysis of Hawking radiation from 2+1 dimensional spinning black hole,} arXiv:0904.2915 [hep-th].

\bibitem{canomaly5}
L. Bonora and M. Cvitan, \emph{Hawking radiation, W-infinity algebra and trace anomalies, JHEP.} {\bf 05} (2008) 071;

R. Becar, P. Gonzalez, G. Pulgar and J. Saavedra, \emph{Anomaly and Hawking radiation from Unruh's and Canonical acoustic black hole,} arXiv:0808.1735 [gr-qc];

L. Bonora, M. Cvitan, S. Pallua and I. Smolic, \emph{Hawking Fluxes, W(infinity) Algebra and Anomalies, JHEP.} {\bf 12} (2008) 021;

V. Akhmedova, T. Pilling, A. de Gill and D. Singleton, \emph{Comments on anomaly versus WKB/tunneling methods for calculating Unruh radiation, Phys. Lett.} {\bf B673} (2009) 227;

A. P. Porfyriadis, \emph{Hawking radiation via anomaly cancelation for the black holes of five-dimensional minimal gauged supergravity, Phys. Rev.} {\bf D79} (2009) 084039; \emph{Anomalies and Hawking fluxes from the black holes of topologically massive gravity, Phys. Lett.} {\bf B675} (2009) 235;

R. Banerjee and B. R. Majhi, \emph{Connecting anomaly and tunneling methods for Hawking effect through chirality, Phys. Rev.} {\bf D79} (2009) 064024;

E. Papantonopoulos and P. Skamagoulis, \emph{Hawking Radiation via Gravitational Anomalies in Non-spherical Topologies, Phys. Rev.} {\bf D79} (2009) 084022;

K. Fujikawa, \emph{Quantum anomalies and some recent developments, Int. J. Mod. Phys.} {\bf A24} (2009) 3306.


\bibitem{canomaly6}
S. Iso, T. Morita and H. Umetsu, \emph{Fluxes of higher-spin currents and Hawking radiations from charged black holes, Phys. Rev.} {\bf D76} (2007) 064015; \emph{Higher-spin Currents and Thermal Flux from Hawking Radiation, Phys. Rev.} {\bf D75} (2007) 124004; \emph{Higher-spin gauge and trace anomalies in two-dimensional backgrounds, Nucl. Phys.} {\bf B799} (2008) 60; \emph{Hawking radiation via higher-spin gauge anomalies, Phys. Rev.} {\bf D77} (2008) 045007.

\bibitem{anomaly1}
R. Banerjee, and S. Kulkarni, \emph{Hawking Radiation and Covariant Anomalies, Phys. Rev.} {\bf D77} (2008) 024018.


\bibitem{anomaly2}
S. Gangopadhyay and S. Kulkarni, \emph{Hawking radiation from Garfinkle-Horowitz-Strominger and nonextremal D1-D5 black holes via covariant anomalies, Phys. Rev.} {\bf D77} (2008) 024038;

R. Banerjee, \emph{Covariant Anomalies, Horizons and Hawking Radiation, Int. J. Mod. Phys.} {\bf D17} (2009) 2539;

S. Gangopadhyay, \emph{Anomalies, horizons and Hawking radiation, Europhys. Lett.} {\bf 85} (2009) 10004; \emph{Hawking radiation from black holes in de Sitter spaces via covariant anomalies,} arXiv:0910.2079 [hep-th];

S. Nam and J. D. Park, \emph{Hawking radiation from covariant anomalies in (2+1)-dimensional black holes, Class. Quant. Grav.} {\bf 26} (2009) 145015;

J. J. Peng and S. Q. Wu, \emph{Covariant anomaly and Hawking radiation from the modified black hole in the rainbow gravity theory, Gen. Rel. Grav.} {\bf 40} (2008) 2619; \emph{Covariant anomalies and Hawking radiation from charged rotating black strings in anti-de Sitter spacetimes, Phys. Lett.} {\bf B661} (2008) 300.


\bibitem{eanomaly}
S. Gangopadhyay, \emph{Hawking radiation from a Reissner-Nordstr?m black hole with a global monopole via covariant anomalies and effective action, Phys. Rev.} {\bf D78} (2008) 044026;

S. Q. Wu and J. J. Peng and Z. Y. Zhao, \emph{Anomalies, effective action and Hawking temperatures of a Schwarzschild black hole in the isotropic coordinates, Class. Quant. Grav.} {\bf 25} (2008) 135001;

Q. Q. Jiang and X. Cai, \emph{Covariant anomalies, effective action and Hawking radiation from Kerr¨CG?del black hole, Phys. Lett.} {\bf B677} (2009) 179.

\bibitem{effective}
R. Banerjee and S. Kulkarni, \emph{Hawking Radiation, Effective Actions and Covariant Boundary Conditions, Phys. Lett.} {\bf B659} (2008) 827;

R. Banerjee and S. Kulkarni, \emph{Hawking Radiation, Covariant Boundary Conditions and Vacuum States, Phys. Rev.} {\bf D79} (2009) 084035;

S. Gangopadhyay, \emph{Hawking radiation from the Garfinkle-Horowitz-Strominger black hole, effective action, and covariant boundary condition, Phys. Rev.} {\bf D77} (2008) 064027.



\bibitem{correction1}
R. Banerjee and B. R, Majhi, \emph{Quantum Tunneling Beyond Semiclassical Approximation, JHEP.} {\bf 0806} (2008) 095;

S. K. Modak, \emph{Corrected entropy of BTZ black hole in tunneling approach, Phys. Lett.} {\bf B671} (2009) 167;

B. R. Majhi, \emph{Fermion tunneling beyond semiclassical approximation, Phys. Rev.} {\bf D79} (2009) 044005;

R. Banerjee, B. R. Majhi and D. Roy, \emph{Corrections to Unruh effect in tunneling formalism and mapping with Hawking effect,} arXiv:0901.0466 [hep-th];

R. Banerjee, B. R. Majhi and E. C. Vagenas, \emph{Quantum tunneling, energy-time uncertainty principle and black hole spectroscopy,} arXiv:0907.4271 [hep-th].



\bibitem{correction2}
J. Y. Zhang, \emph{Black hole quantum tunnelling and black hole entropy correction, Phys. Lett.} {\bf B668} (2008) 353; \emph{Black hole entropy, log correction and inverse area correction, Phys. Lett.} {\bf B675} (2009) 14;

R. G. Cai, L. M. Cao and Y. P. Hu, \emph{Corrected entropy-area relation and modified
Friedmann equations, JHEP.} {\bf 0808} (2008) 090;

T. Zhu, J. R. Ren and M. F. Li, \emph{Influence of generalized and extended uncertainty principle on thermodynamics of FRW universe, Phys. Lett.} {\bf B674} (2009) 204;

T. Zhu and J. R. Ren, \emph{Corrections to Hawking-like radiation for a Friedmann-Robertson-Walker universe, Euro. Phys. J.} {\bf C62} (2009) 413;

T. Zhu, J. R. Ren and M. F. Li, \emph{Corrected entropy of Friedmann-Robertson-Walker universe in tunneling method, JCAP.} {\bf 0908} (2009) 010;

T. Zhu, J. R. Ren and M. F. Li, \emph{Corrected entropy of high dimensional black holes,} arXiv:0906.4194 [hep-th];

M. J. Wang, C.K. Ding and S. B. Chen and J. L. Jing, Gen. Rel. Grav. (in press) (2009).

\bibitem{correction3}
R. Banerjee and B. R. Majhi, \emph{Quantum Tunneling and Trace Anomaly, Phys. Lett.} {\bf B674} (2009) 218;

R. Banerjee and S. K. Modak, \emph{Exact Differential and Corrected Area Law for Stationary Black Holes in Tunneling Method, JHEP.} {\bf 05} (2009) 063.

\bibitem{spectrum}
R. Banerjee and B. R. Majhi, \emph{Hawking black body spectrum from tunneling mechanism, Phys. Lett.} {\bf B675} (2009) 243;

R. Banerjee and S. K. Modak, \emph{Quantum Tunneling, Blackbody Spectrum and Non-Logarithmic Entropy Correction for Lovelock Black Holes, JHEP.} {\bf 0911} (2009) 073.

\bibitem{New1}
C. O. Loust\'{o} and N. S\'{a}nchez, \emph{Back reaction effects in black hole spacetimes, Phys. Lett.} {\bf B212} (1988) 411.

\bibitem{New2}
C. G. Huang, \emph{ Phys. Lett.} {\bf A164} (1992) 384.

\bibitem{other1}
D. V. Fursaev, \emph{Temperature and entropy of a quantum black hole and conformal anomaly, Phys. Rev.} {\bf D51} (1995) 5352;

R. B. Mann and S. N. Solodukhin, \emph{Universality of quantum entropy for extreme black holes, Nucl. Phys.} {\bf B523} (1998) 293;

D. N. Page, \emph{Hawking radiation and black hole thermodynamics, New J. Phys.} {\bf 7} (2005) 203.

\bibitem{other2}
R. K. Kaul and P. Majumdar, \emph{Logarithmic Correction to the Bekenstein-Hawking Entropy, Phys. Rev. Lett.} {\bf 84} (2000) 5255;

S. Kloster, J. Brannlund and A. DeBenedicts, \emph{Phase space and black-hole entropy of higher genus horizons in loop quantum gravity, Class. Quant. Grav.} {\bf 25} (2008) 065008.

\bibitem{other3}
S. Das, P. Majumdar and R. K. Bhaduri, \emph{General logarithmic corrections to black-hole entropy, Class. Quant. Grav.} {\bf 19} (2002)
2355;

S. S. More, \emph{Higher order corrections to black hole entropy, Class. Quant. Grav.} {\bf 22} (2005) 4129;

S. Mukherjee and S. S. Pal, \emph{Logarithmic corrections to black hole entropy and ADS/CFT correspondence, JHEP} {\bf 05} (2002) 026.

\bibitem{other4}
S. Carlip, \emph{Logarithmic corrections to black hole entropy from the Cardy formula, Class. Quant. Grav.} {\bf 17} (2000) 4175;

M. R. Setare, \emph{Logarithmic Correction to the Cardy-Verlinde Formula in Achucarro-Oritz Black Hole, Eur. Phys. J.} {\bf C33} (2004) 555.



\bibitem{New5}
H. Leutwyler, \emph{Gravitational anomalies: A soluble two-dimensional model, Phys. Lett.} {\bf B153} (1985) 65.



\end{thebibliography}
\end{document}